\documentclass[prb,aps,twocolumn,showkeys,showpacs,amsmath,amsfonts,floatfix,superscriptaddress,nofootinbib]{revtex4}
\usepackage{amssymb,amsmath,amstext}                
\usepackage{graphicx}                                               
\usepackage{tikz-network}
\usepackage{epstopdf}                                               
\usepackage{color}
\usepackage{bm}
\usepackage{appendix}                                              
\usepackage[latin2]{inputenc}
\usepackage{ulem}
\usepackage{latexsym}
\usepackage{tabularx}
\usepackage{afterpage}
\usepackage{nonfloat}
\usepackage[colorlinks=true,citecolor=blue,linkcolor=magenta]{hyperref}
\usepackage{lipsum}

\def\be{\begin{equation}}
\def\ee{\end{equation}}
\def\bea{\begin{eqnarray}}
\def\eea{\end{eqnarray}}
\def\bi{\begin{itemize}}
\def\ei{\end{itemize}}

\newcommand{\ket}[1]{\mbox{$| #1 \rangle$}}
\newcommand{\braket}[2]{\mbox{$\langle #1  | #2 \rangle$}}

\begin{document}

\title{ Truncating loopy tensor networks by zero-mode gauge fixing }

\newcommand{\affilju}{
             Jagiellonian University,
             Faculty of Physics, Astronomy and Applied Computer Science,
             Institute of Theoretical Physics,
             ul. \L{}ojasiewicza 11, 30-348 Krak\'ow, Poland
             }

\newcommand{\affildoc}{
             Jagiellonian University,
             Doctoral School of Exact and Natural Sciences,
             ul. \L{}ojasiewicza 11, 30-348 Krak\'ow, Poland
             }

\newcommand{\affilkac}{
             Jagiellonian University,
             Mark Kac Center for Complex Systems Research,
             ul. \L{}ojasiewicza 11, 30-348 Krak\'ow, Poland
             }

\author{Ihor Sokolov}\affiliation{\affilju}\affiliation{\affilkac}
\author{Yintai Zhang}\affiliation{\affildoc}\affiliation{\affilju}
\author{Jacek Dziarmaga}\affiliation{\affilju}\affiliation{\affilkac}

\date{July 28, 2025}

\begin{abstract}
    Loopy tensor networks have internal correlations that often make their compression inefficient. We show that even local bond optimization can make better use of the insight it has locally into relevant loop correlations. By cutting the bond, we define a set of states whose linear dependence can be used to truncate the bond dimension. The linear dependence is eliminated with zero modes of the states' metric tensor.
    The method is illustrated by a series of examples for the infinite pair-entangled projected state (iPEPS) and for the periodic matrix product state (pMPS) that occurs in the tensor renormalization group (TRG) step. In all examples, it provides better initial truncation errors than standard initialization.
\end{abstract}

\maketitle


\section{Introduction}
\label{sec:intro}

Understanding strongly correlated quantum many-body systems is a long-standing problem, especially in two spatial dimensions (2D), where exact diagonalization is limited to small system sizes and quantum Monte Carlo is hampered by the notorious sign problem. The problem is bypassed by tensor networks (TN) that provide an efficient representation for typical ground states of quantum many-body systems~\cite{Verstraete_review_08,Orus_review_14,Nishino_review_2022}. They include the matrix product states (MPS) in one dimension (1D)~\cite{fannes1992,schollwock_review_2011}, the projected entangled pair state (PEPS) in 2D~\cite{nishino01,gendiar03,verstraete2004} and 3D~\cite{Vlaar2021,3D_Charkiv}, and the multi-scale entanglement renormalization ansatz (MERA)~\cite{Vidal_MERA_07,Vidal_MERA_08,Evenbly_branchMERA_14,Evenbly_branchMERAarea_14}.
MPS that is so powerfull in 1D, thanks to its canonical structure, in 2D is limited to small system sizes. This limitation does not apply to PEPS ~\cite{nishino01,gendiar03,verstraete2004, Murg_finitePEPS_07,Cirac_iPEPS_08,Xiang_SU_08,Orus_CTM_09,fu,Lubasch_conditioning,Corboz_varopt_16, Vanderstraeten_varopt_16, Fishman_FPCTM_17, Xie_PEPScontr_17, Corboz_Eextrap_16, Corboz_FCLS_18, Rader_FCLS_18, Rams_xiD_18, Hasik} that is its 2D generalization but, for lack of a canonical form that could be treated efficiently~\cite{canonical_PEPS}, its expressive power may remain largely unused. PEPS has closed loops that makes local optimization of its tensors less effective, as it does not fully account for correlations around the loops. In this paper we show that even local optimization can make better use of the insight it has locally into relevant loop correlations.

\begin{figure}[t!]
\includegraphics[width=0.7\columnwidth]{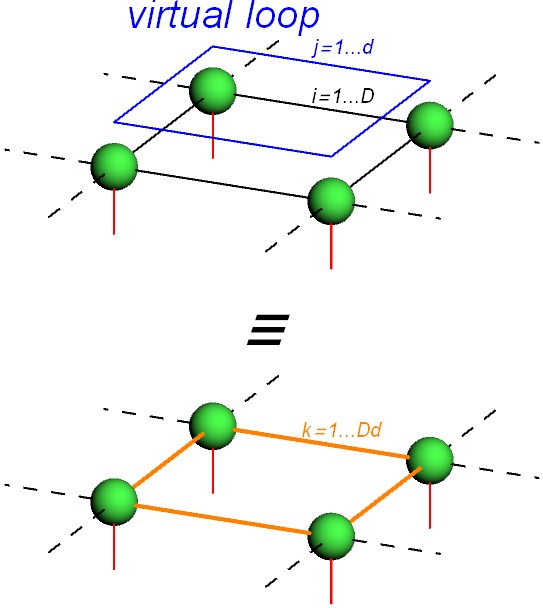}
\caption{
{\bf Cartoon virtual loop entanglement ---}
The top panel shows a four-tensor plaquette in a larger tensor network (TN). The four tensors are contracted by the black bond indices with dimension $D$. Additionally, there is a blue loop carrying a virtual index $j$ that is decoupled from any physical index. The TN state is a sum over $j=1...d$, $\ket{\rm TN}=\sum_{j=1}^d\ket{\psi_j}$, where each state $\ket{\psi_j}$ is the same and proportional to the TN state.
The bottom panel shows the same plaquette after the indices $i$ and $j$ were merged into a single index $k=1...Dd$. Its bond dimension is $d$ times bigger than necessary to represent the TN state. Any single state $\ket{\psi_j}$ with the smaller bond dimension $D$ would suffice to represent the same state: $\ket{\rm TN}\propto\ket{\psi_j}$.
}\label{fig:virtual_loop}
\end{figure}

The nature of the problem can be illustrated by the example in Fig. \ref{fig:virtual_loop}. There is a virtual entanglement loop around a plaquette that is decoupled from physical indices but, nevertheless, parasites bond dimensions along the plaquette's edges. Admittedly, this clear-cut example can probably be fixed in a number of ways but in practice virtual loops are more elusive, as they are neither quite decoupled nor even well defined as loops. Therefore, rather than chasing after a precise definition, we assume a more pragmatic approach. We pick a bond in a tensor network that needs to be truncated, open it for inspection and find an optimal way to truncate its bond dimension by eliminating linear dependence of quantum states constituting the TN. Much of the virtual entanglement is removed along the way.

The paper is organized as follows.
In Sec. \ref{sec:linear} we introduce a simple truncation method, directly inspired by Fig. \ref{fig:virtual_loop}, where the states $\ket{\psi_j}$ are not linearly independent and their linear dependence implies zero modes that can be used to reduce the bond dimension.
In Sec. \ref{sec:toy} a toy example is provided to compare the zero mode truncation with a standard method based on the pseudoinverse.
In Sec. \ref{sec:general} the truncated bond is opened to define a more general form of linear dependence and more general zero modes that allow for more accurate truncation.
In Sec. \ref{sec:eat} we define loopiness of the network, as perceived by the truncated bond, and put the zero-mode truncation (ZMT) in the context of the {\it environment assisted truncation} (EAT) \cite{Hubbard_Sinha}. EAT is the optimal truncation for a bond that perceives the TN as non-loopy.
This Section completes our general considerations, the following Sections are a series of examples.

In Sec. \ref{sec:sudden} we simulate unitary time evolution after a sudden quench in the 2D transverse field quantum Ising model with the iPEPS TN. Even such an unsuspecting set-up can accumulate significant loopiness with evolution time. With the loopiness, the more general ZMT provides a better initial truncation than EAT.
In Sec. \ref{sec:thermal} we consider imaginary time evolution of a purification of the Gibbs thermal state in the Heisenberg model. The purification is represented by iPEPS. The loopiness increases as the temperature is lowered. The initial error after ZMT is found to lie mid way between the initial error after EAT and the final error after the following variational optimization.
In Sec. \ref{sec:Z2} we consider again a sudden quench but this time for the $Z_2$ gauge field, where the plaquette terms in the Kogut-Susskind Hamiltonian generate loop entanglement in the iPEPS TN. EAT, to an even greater extent than a simple initial SVD truncation, fails to manage the entanglement around the plaquettes. They provide poor initial truncation that cannot be compensated by the following variational optimization. ZMT is a better initialization with a better final error after the optimization.
In Sec. \ref{sec:Hubbard} we simulate a sudden quench in the fermionic $t$-$J$ model with a $U(1)\times U(1)$-symmetric iPEPS. ZMT yields an initial truncation error better than mid way between the initial EAT error and the final error after the optimization.
In Sec. \ref{sec:TRG}, we leave the iPEPS realm and consider several initial truncation methods in the tensor renormalization group (TRG) for the classical Ising model. TRG is notoriously plagued by the loop entanglement in a periodic MPS that appears in every coarse-graining step. ZMT is shown to make better initial truncation than the standard one employing the Vidal gauge.
Finally, we conclude in Sec. \ref{sec:concl}.

\begin{figure}[t!]
\includegraphics[width=0.9999\columnwidth]{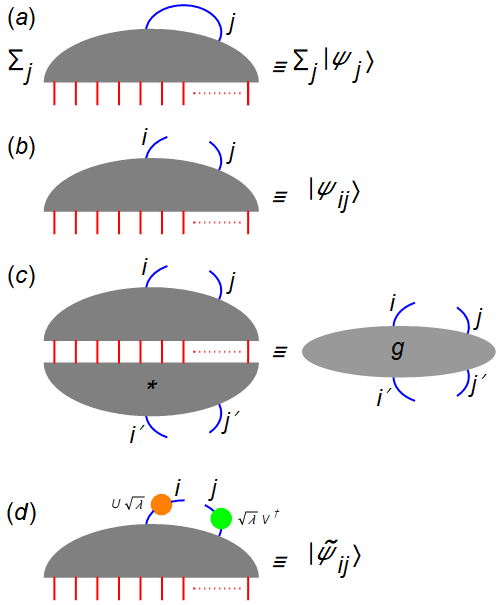}
\caption{{\bf Bond zero modes. ---}
In (a)
the grey semi-ellipsis contains a tensor network (TN). The red lines are its physical indices. All TN's internal bond indices are hidden except for an index $j$ represented by the blue line. Here the summation over $j$ is explicit. Each value of $j$ defines a state $\ket{\psi_j}$.
In (b)
the bond is cut to define more general states $\ket{\psi_{ij}}$. Now the state in (a) can be written as $\ket{\psi}=\sum_{i,j=1}^D\delta_{ij}\ket{\psi_{ij}}$.
In (c)
a contraction of $\ket{\psi_{ij}}$ with its conjugate defines the metric tensor in \eqref{eq:g_ij}.
In (d)
the singular value decomposition \eqref{eq:UlambdaV} is absorbed into the TN to define the new states in \eqref{eq:tilde_psi}.
In (e)
a conventional gauge transformation, inserting $S^{-1}S$ in the bond, defines the new states in \eqref{eq:psi^S}.
}
\label{fig:psi_j}
\end{figure}

\section{Elimination of linear dependence}
\label{sec:linear}

Figure \ref{fig:psi_j} (a) shows a black box containing a TN state $\ket{\psi}$ that demonstrates to the outside world only its physical indices. The internal network of local tensors contracted through bond indices is not shown except for an explicit summation over one bond index: $j=1,...,D$. Each value of $j$ defines a state $\ket{\psi_j}$. The TN state is a sum of these states: $\ket{\psi}=\sum_{j=1}^D \ket{\psi_j}$. Their linear (in)dependence can be verified by diagonalizing the matrix of their overlaps:
\be
g_{ij} \equiv \braket{\psi_i}{\psi_j} = \sum_{k=1}^D U_{ik} N_k U^*_{jk},
\label{eq:g_i}
\ee
where $N_1\geq\dots\geq N_D$. When there is a zero mode $Z_j\equiv U_{jD}$ with $N_D=0$ then they are linearly dependent:
\be
\sum_{j=1}^D Z_j \ket{\psi_j} = 0.
\label{eq:linear_dep}
\ee
The order of indices $j$ can be permuted to make $|Z_j|$ maximal for $j=D$. With \eqref{eq:linear_dep} one of the states $\ket{\psi_j}$ can be eliminated from the TN state. Towards this end, we note that there is a zero-mode gauge freedom to choose a parameter $z$:
\be
\sum_{j=1}^D \ket{\psi_j} =
\sum_{j=1}^{D} \left( 1 + z Z_j \right) \ket{\psi_j}.
\label{eq:psi_z}
\ee
Fixing the gauge as $z=-1/Z_D$ eliminates $\ket{\psi_D}$:
\be
\sum_{j=1}^D \ket{\psi_j} =
\sum_{j=1}^{D-1} \left(1-Z_j/Z_D\right) \ket{\psi_j} =
\sum_{j=1}^{D-1} \ket{\tilde\psi_j}.
\label{eq:overlap_truncation}
\ee
In the last step $1-Z_j/Z_D$ was absorbed into tensors contracted through the index $j$ in Fig. \ref{fig:psi_j} (a). The elimination truncated the bond dimension from $D$ to $D-1$.

Suppose now that the lowest eigenmode has a small but non-zero eigenvalue, $N_D>0$. To leading order in $N_D$ we can still use it as if it were a zero mode, see App. \ref{app:imperfect_ld}. The approximate elimination changes the state by $-\sum_{j=1}^{D  } \left(Z_j/Z_D\right) \ket{\psi_j}$. The norm squared of this change is
\be
f=
\frac{ \sum_{i,j=1}^D Z^*_{i} g_{ij} Z_{j} }
     { \left|         Z_D       \right|^2 } =
\frac{                 N_D                   }
     { \left|         Z_D      \right|^2 }.
\label{eq:fD}
\ee
The elimination of the $\ket{\psi_D}$ with the maximal $|Z_D|$ results in the most accurate truncation of the bond dimension. Furthermore, the lowest eigenmode may turn out not to be the optimal one as it does not need to have the lowest $f$, where not only $N_D$ but also $Z_D$ matters. 

Further investigation may be required to see a possible role of degeneracies/multiplets of the lowest eigenvalues $N_k$. A similar issue was shown to be very important in the traditional truncation by the singular value decomposition~\cite{Hasik}.

\section{Zero-mode gauge fixing versus \\
            pseudoinverse approach}
\label{sec:toy}

The zero mode elimination can be compared with a standard truncation. To this end, let us consider a simple example where the target TN state is $\frac12\sum_{j=1}^2\ket{\psi_j}$ and the two normalized states are actually the same: $\ket{\psi_1}=\ket{\psi_2}$. We want to make a variational state $\ket{\psi}=\sum_{j=1}^2 c_j \ket{\psi_j}$ as close to the target state as possible. The norm squared of their difference is
\be
f = c^\dag g ~c - (1,1) c - c^\dag (1,1)^T + 1.
\ee
Here $c=(c_1,c_2)^T$ is a vector of variational parameters and $g=1+\sigma^x$ is singular with a zero mode $(1,-1)^T$. The norm is minimal when $c$ satisfies a linear equation
\be
g ~ c = (1,1)^T.
\ee
It is standard to solve it by the pseudoinverse as
\be
c = ~{\rm pinv}(g) ~ (1,1)^T = \frac12 (1,1)^T
\ee
but this solution yields the original TN with bond dimension two: $\ket{\psi}=\frac12\sum_{j=1}^2\ket{\psi_j}$. However, a general solution with $f=0$ is
\be
c = \frac12 (1,1)^T + z (1,-1)^T,
\ee
where $z$ is a free parameter. Employing this zero-mode gauge freedom, we can set $z=\frac12$ to make the optimal variational state have bond dimension one: $\ket{\psi}=\ket{\psi_1}$.

\section{General bond zero modes}
\label{sec:general}

In Fig. \ref{fig:psi_j} (b) we define more general states $\ket{\psi_{ij}}$. In their terms, the TN state in Fig. \ref{fig:psi_j} (a) becomes
\be
\ket{\psi}=\sum_{i,j=1}^D\delta_{ij}\ket{\psi_{ij}}.
\label{eq:psi_ij}
\ee
A more general metric tensor is
\be
g_{ij,i'j'}=\braket{\psi_{ij}}{\psi_{i'j'}},
\label{eq:g_ij}
\ee
see Fig. \ref{fig:psi_j} (c). Suppose that it has a zero mode $Z$, satisfying
$
g_{ij,i'j'} Z_{i'j'} = 0,
$
normalized as ${\rm Tr}~Z^\dag Z=1$. It provides a gauge freedom to rewrite \eqref{eq:psi_ij} as
\be
\ket{\psi}=
\sum_{i,j=1}^D
\left(
\delta_{ij} + z ~ Z_{ij} \right)
\ket{\psi_{ij}}
\label{eq:psi_z_gen}
\ee
with an arbitrary parameter $z$. The freedom can be used to truncate the bond dimension by a suitable choice of $z$.

When the matrix form of the zero mode is diagonalizable, then
\be
Z_{ij} = \sum_{k=1}^D S^{-1}_{ik} E_k S_{kj}
\label{eq:Zdiag}
\ee
with eigenvalues ordered as $|E_1|\leq\dots\leq|E_D|$. For any $k$ the choice $z=-1/E_k$ makes the matrix $\delta_{ij} + z ~ Z_{ij}$ singular. We choose $z=-1/E_D$ for the sake of numerical stability. The singular value decomposition
\be
\delta_{ij} - ~ Z_{ij}/E_D = \sum_{k=1}^D U_{ik} \lambda_k V^*_{jk},
\label{eq:UlambdaV}
\ee
has a zero singular value, $\lambda_D=0$. The same conclusion holds also when the similarity transformation \eqref{eq:Zdiag} brings $Z_{ij}$ to a general Jordan form.

Truncating $\lambda_D=0$, we can rewrite \eqref{eq:psi_z_gen} as
\be
\ket{\psi} = \sum_{k=1}^{D-1} \lambda_k \sum_{i,j=1}^D U_{ik} V^*_{jk} \ket{\psi_{ij}}.
\ee
By absorbing the unitary matrices $U$ and $V$ into the two local tensors of $\ket{\psi_{i'j'}}$, as in Fig. \ref{fig:psi_j} (d), we can define new states:
\be
\ket{\psi_{ij}^\lambda} =
\sqrt{\lambda_i\lambda_j}
\sum_{i',j'=1}^D U_{i'i} V^*_{j'j} \ket{\psi_{i'j'}}.
\label{eq:tilde_psi}
\ee
In their terms the TN state becomes
\be
\ket{\psi}=
\sum_{i,j=1}^{D-1}
\delta_{ij}
\ket{\psi_{ij}^\lambda}.
\label{eq:tilde_psi_ij}
\ee
When compared to the original state \eqref{eq:psi_ij}, its bond dimension got compressed from $D$ to $D-1$.

From a more general perspective, the elimination of linear dependence in Sec. \ref{sec:linear} is a special case of the more general zero-mode gauge fixing. In Sec. \ref{sec:linear} the metric tensor $g_{ij,i'j'}$ is reduced to the diagonal subspace where $i=j,i'=j'$ and $Z_{ij}=\delta_{ij}Z_j$.
Furthermore, when the zero mode $Z_{ij}$ is diagonalizable as in \eqref{eq:Zdiag}, the generalized zero mode can be brought to the diagonal one by inserting a conventional gauge transformation in the bond, $1=S^{-1}S$. The transformation defines new states
\bea
\ket{\psi_{ij}^S} & = &
\sum_{i',j'=1}^D S^{-1}_{i'i} S_{jj'} \ket{\psi_{i'j'}},
\label{eq:psi^S}
\eea
see Fig. \ref{fig:psi_j} (e), and in their representation the zero mode becomes $Z^S_{ij}=\delta_{ij}E_j$.
In a similar vein to the diagonal reduction, in the following we consider also other subspaces: a Hermitian $Z_{ij}=Z^*_{ji}$ for a complex TN and a real symmetric $Z_{ij}=Z_{ji}$ for a real one.

The general zero-mode gauge fixing remains useful for truncation even when there are no exact zero modes. The state \eqref{eq:psi_z_gen} remains applicable as a variational ansatz even when the lowest eigenvalue of $g_{ij,i'j'}$ is nonzero, $N>0$, and the zero-mode gauge symmetry is only approximate. To the leading order in small $N$, we can keep $z=-1/E_D$ as for a zero mode, but now the norm squared of the difference between the truncated and the target state is equal to
\be
f = \frac{N}{|E_D|^2},
\label{eq:fDDmin}
\ee
see App. \ref{app:imperfect}. As anticipated, it is optimal to choose $z=-1/E_D$ with the largest-magnitude eigenvalue $E_D$, but it is not necessarily optimal to choose the eigenmode with the lowest $N$. The optimal eigenmode is the one with the lowest $f$, as both $N$ and its maximal $|E_D|$ matter.
A comparison between \eqref{eq:fD} and \eqref{eq:fDDmin} helps to determine when to switch from the cheaper linear elimination to the more general approximate zero modes. The former involves the smaller $D\times D$ metric tensor $g_{ij}$ while the latter the bigger $D^2\times D^2$ metric $g_{ii',jj'}$.

Finally, in App. \ref{app:beyond} we consider a perturbative modification of the optimal eigenmode from $Z_{ij}$ to ${\cal Z}_{ij}$ with a lower truncation error $f$ in \eqref{eq:fDDmin}, but the improvement $\propto f^2$ may not justify the numerical overhead.

\begin{figure}[t!]
\includegraphics[width=0.9999\columnwidth]{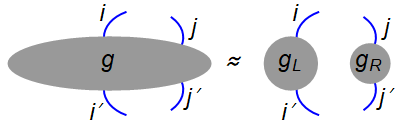}
\caption{{\bf Environment assisted truncation (EAT). ---}
The metric tensor in Fig. \ref{fig:psi_j} (c) is singular value decomposed between its left and right indices and then, in \eqref{eq:g_eat}, approximated by truncation to the leading singular value. The tensors $g_{L,R}$ can be made Hermitian and non-negative. By definition, a tensor network appears non-loopy to the bond when this approximation is exact.
}
\label{fig:gLgR}
\end{figure}

\section{Loopy versus non-loopy metric tensor}
\label{sec:eat}

The metric tensors, $g_{ij}$ and $g_{ii',jj'}$, depend on the gauge of the considered bond and so does the zero mode truncation. Ref. \onlinecite{Hubbard_Sinha} introduced a particularly suitable gauge for {\it environment assisted truncation} (EAT). The EAT gauge can be fixed as follows. It is similar the gauge introduced in Ref. \onlinecite{Evenbly_loops}.

The metric tensor $g_{ij,i'j'}$ can be singular value decomposed between its left ($ii'$) and right ($jj'$) indices and then approximated by keeping just the leading singular value $\lambda_1$ as
\be
g_{ij,i'j'} \approx g_L^{ii'} ~\lambda_1~ g_R^{jj'},
\label{eq:g_eat}
\ee
where $g_{L,R}$ can be made Hermitian and non-negative, see Fig. \ref{fig:gLgR}. It is the best product approximation to the full metric tensor with respect to the Frobenius norm. When the considered bond is the only connection between the left and the right part of the TN then it is exact. In this sense we can measure loopiness of the TN, as perceived by the bond, by a ratio
\be
l = \frac{\lambda_2}{\lambda_1},
\label{eq:l}
\ee
where $\lambda_2$ is the second largest singular value. When there are no loops affecting the bond, then there is just one non-zero singular value and the product \eqref{eq:g_eat} is exact.

In order to fix the EAT gauge, first the left and right metric tensors are diagonalized as
\be
g_L = U_L N_L U_L^\dag, ~~
g_R = U_R N_R U_R^\dag.
\ee
Then a singular value decomposition,
\be
N_L^{1/2} ~ U_L^T ~ U_R ~ N_R^{1/2} = W_L ~\Lambda~ W_R,
\ee
where $W_{L,R}$ are unitary and $\Lambda$ is a diagonal matrix of singular values, defines the EAT gauge transformation,
\be
1~=~
\left(
U_L^* \mu_L^{-1/2} W_L \Lambda^{1/2}
\right)
\left(
\Lambda^{1/2} W_R \mu_R^{-1/2} U_R^\dag
\right),
\label{eq:eat-gauge}
\ee
to be inserted in the considered bond. When \eqref{eq:g_eat} is exact then the transformation defines the Schmidt decomposition between the left and right parts of the TN with the entanglement spectrum $\Lambda$. This motivates the environment assisted truncation (EAT) \cite{Hubbard_Sinha}, where $\Lambda$'s are truncated to initialize variational tensors.
With or without the truncation, in the EAT gauge \eqref{eq:eat-gauge} the left and right metric tensors are diagonal:
$
\tilde{g}^L = \Lambda = \tilde{g}^R,
$
but in general the full metric tensor is not their product: $\tilde{g}\neq \Lambda\otimes\Lambda$. The inequality indicates non-trivial loopiness of the network as perceived by the bond.

In the non-loopy case, when $\tilde{g}=\Lambda\otimes\Lambda$, the truncation by the zero-mode gauge fixing is the same as EAT. Indeed, the lowest eigenmode of $\tilde g$ is $\ket{Z}=\ket{D}\ket{D}$ with the eigenvalue $N$ equal to the lowest $\Lambda_D$. Using it for the zero mode truncation, as in Sec. \ref{sec:general}, is equivalent to truncating the lowest $\Lambda_D$ that is in turn equivalent to EAT. Therefore, in the non-loopy case, the zero-mode gauge fixing in the EAT gauge is optimal, just as is EAT. However, for a general loopy bond metric tensor the zero-mode gauge fixing is capable of truncating a loop while EAT is not.

\begin{figure}[t!]
\includegraphics[width=0.9999\columnwidth]{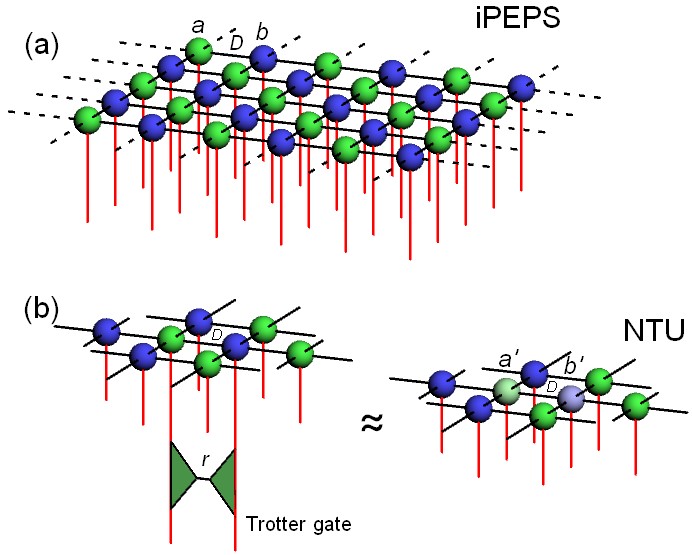}
\caption{
{\bf Neighborhood tensor update (NTU). }
In (a) the infinite PEPS (iPEPS) tensor network with two sublattice tensors $a$ and $b$.
In (b) left, a two-site Trotter gate is applied to a pair of tensors. The gate's rank is $r$ and its application increases the bond dimension from $D$ to $rD$.
In (b) right, the dimension is truncated back to $D$. The initial error of the truncation is the Frobenius norm of the difference between the left and the right. After the initialization, the two tensors $a'$ and $b'$ on the right are further optimized variationally to minimize the error.
}\label{fig:ntu}
\end{figure}

\begin{figure}[t!]
\includegraphics[width=0.9999\columnwidth]{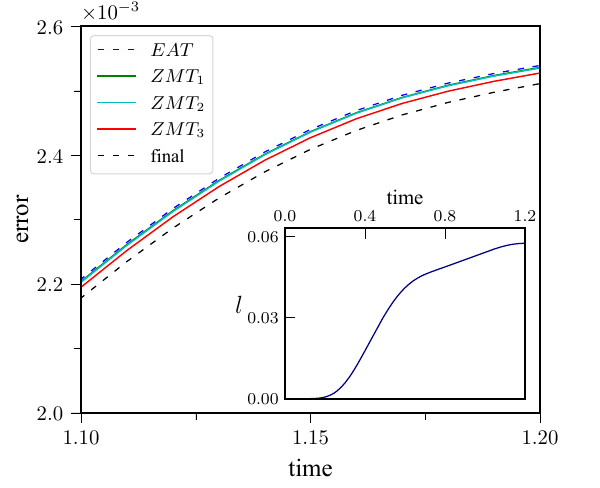}
\caption{
{\bf Quantum Ising model - sudden quench. }
The initial truncation errors and the final error in function of time after the sudden quench.
The inset shows loopiness in function of time.
Here the iPEPS bond dimension $D=8$ and the time step $dt=0.01$.
}\label{fig:sudden}
\end{figure}

\section{Sudden quench in the quantum ising model}
\label{sec:sudden}

We consider the genuinely 2D PEPS tensor network \cite{nishino01, gendiar03, verstraete2004, Murg_finitePEPS_07,Cirac_iPEPS_08,Xiang_SU_08,Gu_TERG_08,Orus_CTM_09,Lubasch_conditioning,
fu,Corboz_varopt_16, Vanderstraeten_varopt_16, Fishman_FPCTM_17, Xie_PEPScontr_17, Corboz_Eextrap_16, Corboz_FCLS_18, Rader_FCLS_18, Rams_xiD_18} shown in Fig. \ref{fig:ntu} (a). The TN has closed loops in distinction to the 1D MPS. The infinite PEPS ansatz (iPEPS) was used to simulate unitary time evolution after a sudden Hamiltonian quench \cite{CzarnikDziarmagaCorboz,HubigCirac,tJholeHubig,Abendschein08,SUlocalization,SUtimecrystal,ntu,mbl_ntu,BH2Dcorrelationspreading,ising2D_correlationsperading,schmitt2021quantum,Mazur_BH,Corboz_SF}. Given PEPS's non-canonical structure, it seems necessary to resort to local updates in time evolution, like the neighborhood tensor update (NTU) \cite{ntu}, see Fig. \ref{fig:ntu} (b), that was used previously to simulate the many-body localization \cite{mbl_ntu}, the Kibble-Zurek ramp in the Ising and Bose-Hubbard models \cite{schmitt2021quantum,Mazur_BH,Science_Dwave}, thermal states obtained by imaginary time evolution in the fermionic Hubbard model \cite{Hubbard_Sinha,Sinha_Wietek_Hubbard}, as well as bang-bang preparation of quantum states \cite{BB_iPEPS,BB_iPEPS_finite}.

Here we reconsider a sudden quench in the quantum Ising model on an infinite 2D square lattice:
\be
H = -\sum_{\langle i,j \rangle} \sigma_i^z \sigma_j^z - g \sum_j \sigma_j^x.
\label{eq:Hising}
\ee
The system is initialized in a state fully polarized along $+x$ and then, at the time $t=0$, evolved with the critical transverse field $g_c=3.04438$ \cite{Deng_QIshc_02}. After every two-site Trotter gate the bond dimension doubles and needs to be truncated back. The truncation is done by the neighborhood tensor update~\cite{ntu}, see Fig. \ref{fig:ntu} (b). After the initial truncation of the two tensors affected by the gate, they are optimized variationally until the error of the truncation is minimized.

In Fig. \ref{fig:sudden} we compare the initial errors after different truncation schemes. We also plot the loopiness \eqref{eq:l} in function of time.
In the simple Ising model the final error does not depend on the initial truncation scheme, but the initial error does.
The basic scheme is EAT in Sec. \ref{sec:eat} that ignores the non-zero loopiness.
In ZMT$_1$ scheme we fix the EAT gauge and then use the metric $g_{ij}=g_{ii,jj}$ in \eqref{eq:g_ij} and truncate as in \eqref{eq:overlap_truncation}. The dimension is cut one-by-one down to $D$.
In ZMT$_2$ scheme we fix the EAT gauge and then cut the dimension one-by-one as in Sec. \ref{sec:general} but with eigenmodes restricted to the Hermitian subspace, $Z_{ij}=Z_{ji}^*$, where eigenvalues $E_k$ are real and eigenvectors $S_{kj}$ orthonormal.
The third scheme, ZMT$_3$, is the same as ZMT$_2$ but without any restrictions on $Z_{ij}$ and without the initial EAT gauge fixing. It literally follows Sec. \ref{sec:general}.
The results in Fig. \ref{fig:sudden} show that ZMT$_{1,2}$ are slightly better than EAT but ZMT$_3$ is better than ZMT$_{1,2}$ and its initial truncation error lies somewhat less than mid way between the initial error of EAT and the final error after the following variational optimization. In the simple example the final error does not depend on the initial truncation method.

\begin{figure}[t!]
\includegraphics[width=0.9999\columnwidth]{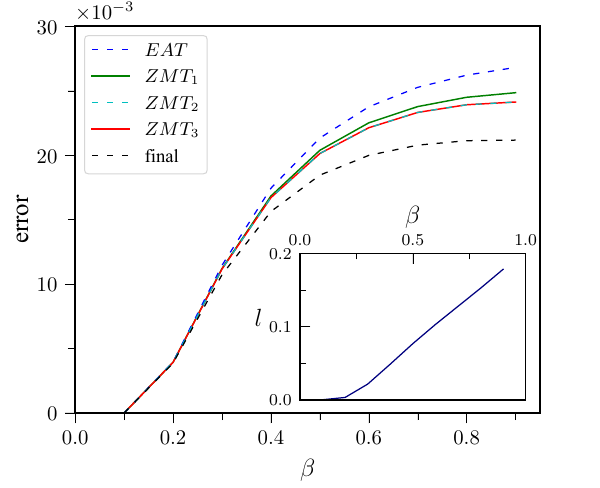}
\caption{
{\bf Heisenberg model - thermal state. }
The initial truncation errors and the final error in function of the inverse temperature $\beta$.
The inset shows loopiness in function of $\beta$.
Here the iPEPS bond dimension $D=5$ and the imaginary time step $d\beta=0.1$.
}\label{fig:thermal}
\end{figure}

\section{Thermal state in the Heisenberg model}
\label{sec:thermal}

The iPEPS can also be used to represent purification of the Gibbs state \cite{CzarnikDziarmagaCorboz,ntu,Hubbard_Sinha,Sinha_Wietek_Hubbard}. Here we consider the Heisenberg model on an infinite square lattice:
\be
H =
\sum_{\langle i,j \rangle}
\left(
\sigma^x_i \sigma^x_j +
\sigma^y_i \sigma^y_j +
\sigma^z_i \sigma^z_j
\right).
\ee
The thermal state in function of $\beta$ is obtained by imaginary time evolution of the purification in the Hermitian gauge as in Ref. \onlinecite{ntu}.

In Fig. \ref{fig:thermal} we compare the initial errors after different truncation schemes. We also plot the loopiness in function of $\beta$.
The basic scheme is EAT that ignores the loopiness altogether.
In ZMT$_1$ scheme we fix the EAT gauge and then use the metric $g_{ij}=g_{ii,jj}$ in \eqref{eq:g_ij} and truncate as in \eqref{eq:overlap_truncation}. The dimension is cut one-by-one down to $D$.
In ZMT$_2$ scheme we fix the EAT gauge and then cut the dimension one-by-one as in Sec. \ref{sec:general} but with eigenmodes restricted to the real-symmetric subspace, $Z_{ij}=Z_{ji}$, where eigenvalues $E_k$ are real and eigenvectors $S_{kj}$ orthonormal.
The third scheme, ZMT$_3$, is the same as ZMT$_2$ but without any restrictions on $Z_{ij}$ and without the initial EAT gauge fixing, but $E_D$ is selected as the largest magnitude real eigenvalue of $Z_{ij}$.
The data in Fig. \ref{fig:thermal} show improvement between EAT and ZMT$_1$ and then between ZMT$_1$ and ZMT$_{2,3}$, where ZMT$_{2,3}$ yield similar errors.
The initial errors of the best schemes ZMT$_{2,3}$ lie mid way between the initial error after EAT and the final error after the variational optimization.

\begin{figure}[t!]
\includegraphics[width=0.9999\columnwidth]{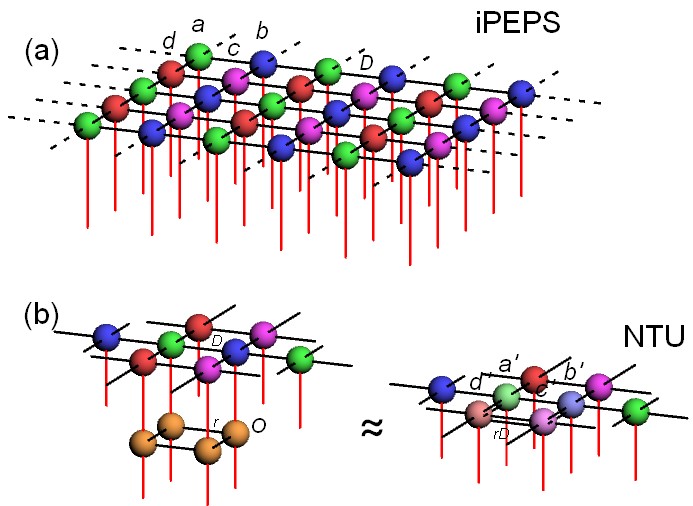}
\caption{
{\bf NTU for gauge field. }
In (a) the infinite PEPS (iPEPS) tensor network ansatz with four sublattice tensors: $a,b,c,d$. The dimension of all bond indices is $D$.
In (b) left, the plaquette evolution operator \eqref{eq:U_p}, in the matrix product operator (MPO) representation \eqref{eq:MPO}, is applied to the $abcd$ plaquette of iPEPS tensors. Here the MPO bond dimension $r=2$.
In (b) right, the central bond $a'-b'$ is truncated to $D$ and then the truncated tensors $a',b'$ are optimized to minimize the difference between the target left and the variational right diagram.
After the minimization is completed, the next bond $b'-c'$ is zero-mode truncated and optimized in a similar tensor neighbourhood centered on the bond.
Then the same procedure is repeated for the bonds $c'-d'$ and $d'-a'$.
}\label{fig:ntu_gauge}
\end{figure}

\begin{figure}[t!]
\includegraphics[width=0.9999\columnwidth]{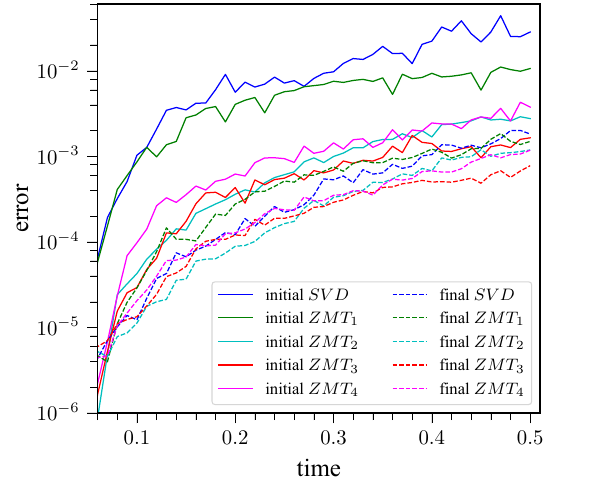}
\caption{
{\bf $Z_2$ gauge field - sudden quench. }
Truncation errors averaged over the four bonds in Fig. \ref{fig:ntu_gauge}: $a'-b'$, $b'-c'$, $c'-d'$, and $d'-a'$. Solid lines are the initial errors right after the initial truncation and the dashed lines are final final errors after the following optimization.
Here $D=10$, $g=1$, and $dt=0.01$.}\label{fig:Z2_results_1}
\end{figure}

\begin{figure}[t!]
\includegraphics[width=0.9999\columnwidth]{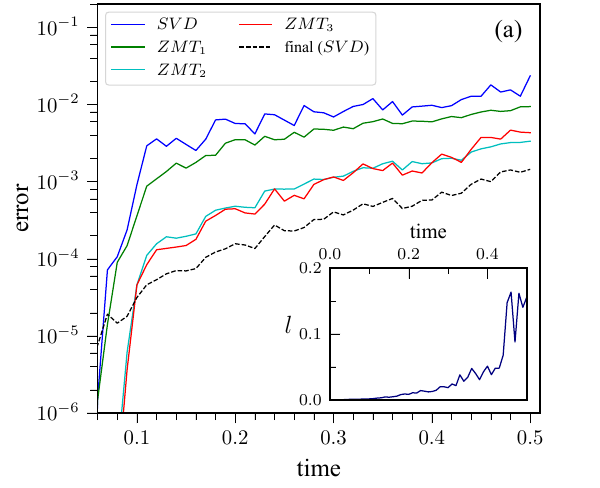}
\includegraphics[width=0.9999\columnwidth]{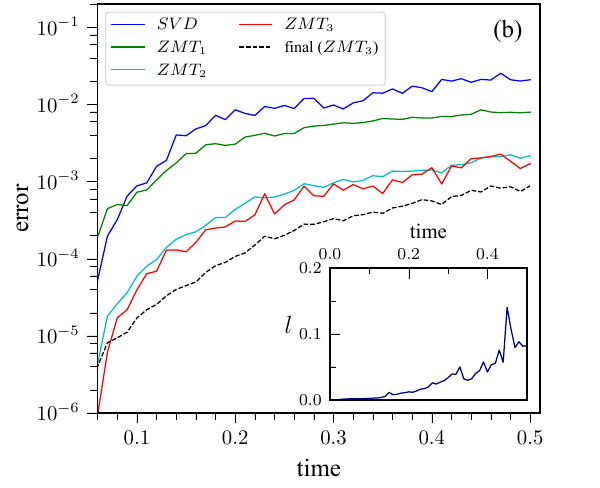}
\caption{
{\bf $Z_2$ gauge field - sudden quench. }
The same time evolution as in Fig. \ref{fig:Z2_results_1}, but here either SVD (top) or ZMT$_3$ (bottom) is used for the initial truncation and then followed by the optimization. The initial errors of the other truncation methods are calculated for comparison only. Their truncated tensors are not used in the following evolution and discarded.
Here $D=10$, $g=1$, and $dt=0.01$.
}\label{fig:Z2_results_2}
\end{figure}

\begin{figure}[t!]
\includegraphics[width=0.9999\columnwidth]{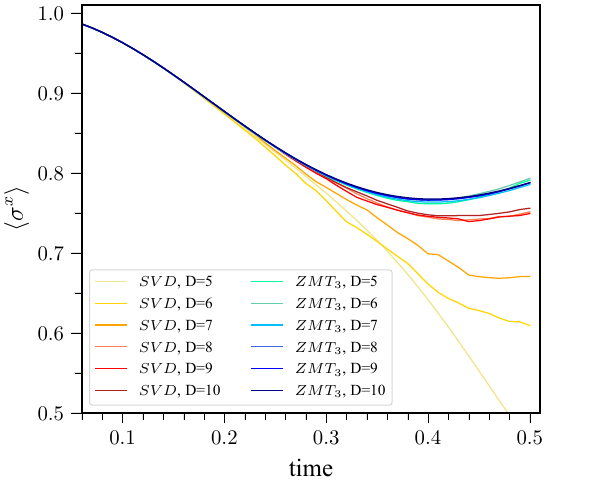}
\caption{
{\bf $Z_2$ gauge field - sudden quench. }
The transverse magnetization $\langle\sigma^x\rangle$ in the time evolution in Fig. \ref{fig:Z2_results_1} with either SVD or ZMT$_3$ initial truncation followed by the optimization. ZMT$_3$ quickly converges with increasing $D$ while SVD is not converged up to $D=10$. 
Here $g=1$ and $dt=0.01$.
}\label{fig:Z2_results_X}
\end{figure}

\section{$Z_2$ gauge field - sudden quench}
\label{sec:Z2}

The $Z_2$ gauge field Hamiltonian on an infinite square lattice is
\be
H = H_m + H_e \equiv -\sum_p \sigma^z_{p_1} \sigma^z_{p_2} \sigma^z_{p_3} \sigma^z_{p_4} - g \sum_s \sigma^x_s.
\ee
In the magnetic Hamiltonian $H_m$, index $p$ runs over white $2\times 2$ plaquettes on a checkerboard plaquette tiling. The indices $p_i$ number the $p$-th plaquette's four corner sites. In the electric Hamiltonian $H_e$, index $s$ runs over lattice sites, and $g$ is the strength of the electric term. Here we employ a dual lattice, where fermionic fields would live on sites in the center of every black plaquette. In the absence of any charges, the Gauss law requires products $\sigma^x_{p_1}\sigma^x_{p_2}\sigma^x_{p_3}\sigma^x_{p_4}=1$ on every black plaquette $p$.

A charge-less product state with all spins polarized along $+\sigma^x$ is prepared and then evolved in time with the total $H$. We simulate the evolution representing the state by an iPEPS with four sublattice tensors, see Fig. \ref{fig:ntu_gauge} (a). Every discrete time step $dt$ is subject to the second order Suzuki-Trotter decomposition:
\be
e^{-idt H}
\approx
e^{-\frac12 idt H_e}
e^{-idt H_m}
e^{-\frac12 idt H_e}.
\ee
The electric evolution operator is a product of local site operators $e^{-\frac12 idt H_e} = \prod_s e^{\frac12 i gdt \sigma^x_s}$. Each iPEPS tensor is simply subject to the local transformation $e^{\frac12 i gdt \sigma^x_s}$. The magnetic evolution operator $e^{-idt H_m}$ is a product of commuting operators,
\bea
& &
U_p(dt) = \nonumber\\
& &
e^{i dt~ \sigma^z_{p_1} \sigma^z_{p_2} \sigma^z_{p_3} \sigma^z_{p_4}} = \nonumber\\
& &
1_{p_1} 1_{p_2} 1_{p_3} 1_{p_4} ~\cos dt~ +
\sigma^z_{p_1} \sigma^z_{p_2} \sigma^z_{p_3} \sigma^z_{p_4} ~i\sin dt,
\label{eq:U_p}
\eea
applied to the white plaquettes $p$. Here we have a sum of two terms. In the first each tensor on the plaquette is applied with $(\cos dt)^{1/4} 1_{p_i}\equiv O_{p_i}^{1}(dt)$ and in the second with $(i \sin dt)^{1/4} \sigma^z_{p_i}\equiv O_{p_i}^{2}(dt)$. The magnetic evolution operator can then be represented by a periodic matrix product operator (pMPO):
\be
U_p(dt) = \sum_{j=1,2} O_{p_1}^{j}(dt) O_{p_2}^{j}(dt) O_{p_3}^{j}(dt) O_{p_4}^{j}(dt).
\label{eq:MPO}
\ee
Here $j$ is a virtual loop index around the plaquette, see Fig. \ref{fig:ntu_gauge} (b). In every time step the commuting pMPOs are applied to all $a-b-c-d-$ plaquettes first, as in Fig. \ref{fig:ntu_gauge} (b), and then to all $c-d-a-b-$ plaquettes.
Their repeated application in every time step can proliferate virtual loop entanglement.

The loop entanglement makes EAT fail. EAT obliterates the effect of the pMPO gate: the SVD$_1$ in Fig. \ref{fig:gLgR} truncates its effect in a way that cannot be corrected by the following optimization. This is why we use as a benchmark a simple SVD truncation between the two tensors on the truncated bond (SVD). The other methods include ZMT$_1$, eliminating the linear dependence as in Sec. \ref{sec:linear}, ZMT$_2$ with a Hermitian $Z_{ij}=Z^*_{ji}$, and ZMT$_3$ as in Sec. \ref{sec:general}.
Additionally, we include ZMT$_4$, where we assume a product ansatz $Z_{ij}=R_iL_j$ with $R^\dag R=L^\dag L=1$. The vectors are optimized iteratively, $\to L\to R\to$, until convergence of the cost function \eqref{eq:fDDmin}, where $N=Z^\dag g Z$ is an expectation value and $E_D=\sum_j L_j R_j$.
The results are collected in Figs. \ref{fig:Z2_results_1} and \ref{fig:Z2_results_2}. In Fig. \ref{fig:Z2_results_1} the same evolution is simulated with several initial truncations. We show both the initial and the final error after the optimization. Overall, it is the general ZMT$_3$ that provides the best initial and final errors.
In order to provide a slightly different perspective, in Fig. \ref{fig:Z2_results_2} we run the SVD-truncated and the ZMT$_3$-truncated evolution but at every gate we also try other truncation schemes just to compare their initial truncation errors (but without any further use of their truncated tensors).

Unlike the quantum Ising and Heisenberg models, the gauge field Hamiltonian proves complex enough computationally to make the final truncation error, after the optimization, depend on the initialization.
This affects observables, as shown in Fig. \ref{fig:Z2_results_X}, where the transverse magnetization quickly converges with $D$ after the ZMT$_3$-initialization in sharp contrast to the SVD-initialization.

\begin{figure}[t!]
\includegraphics[width=0.9999\columnwidth]{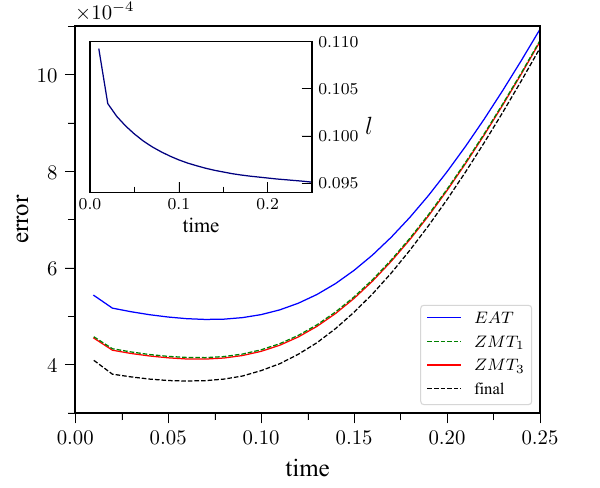}
\caption{
{\bf $t$-$J$ model - sudden quench. }
The initial ground state at half-filling is suddenly depleted of $25\%$ particles and then evolved with the $t$-$J$ Hamiltonian. The final error does not depend on the initialization. The initial error is the best for ZMT$_{1,3}$ and much better than the initial EAT error. The loopiness is decreasing with time.
Here $t=1$, $J=0.5$, $D=10$, and the time step $dt=0.01$. $\textrm{ZMT}_{1,3}$ are initialized with the EAT gauge.
}\label{fig:tJ}
\end{figure}

\section{Sudden quench in the $t$-$J$ model}
\label{sec:Hubbard}

One of the simplest models of interacting fermions on a lattice is the $t$-$J$ Hamiltonian on a square lattice \cite{Chao_1977},
\begin{align}
H \;=\;
& -t \sum_{\langle i,j \rangle,\sigma}
\Bigl(
\tilde c_{i\sigma}^\dagger \,\tilde c_{j\sigma} + \tilde c_{j\sigma}^\dagger \,\tilde c_{i\sigma}
\Bigr) \nonumber\\[2mm]
&+\; J \sum_{\langle i,j\rangle}
\Bigl(
\mathbf{S}_i \cdot \mathbf{S}_j - \tfrac{n_i n_j}{4}
\Bigr).
\label{eq:tJ-Ham}
\end{align}
Here the operator $\tilde c_{i\sigma}$ is an annihilation operator for a fermion with spin $\sigma=\uparrow,\downarrow$ at site $i$ projecting out double occupancy, and $n_i$ is the number operator. We use the NTU algorithm of Ref.~\cite{Hubbard_Sinha,YASTN1} for a $U(1)\times U(1)$-symmetric iPEPS, enriched with the zero mode gauge fixing, to simulate a sudden quench.

First, we prepare the ground state at half-filling, when the Hamiltonian is equivalent to the Heisenberg model, using the variational optimization of iPEPS \cite{Liao2019,Hasik2021} with bond dimension $D_0=7$ on a checkerboard lattice. The ground state has the anti-ferromagnetic N\'eel order. We define a $2\times2$ unit cell and apply $\tilde c_{i\uparrow}$ at the top-left site of every unit cell of the infinite lattice. After the annihilation, we let the system evolve with the Hamiltonian \eqref{eq:tJ-Ham}. In Fig. \ref{fig:tJ} we follow the average of the truncation errors on all 4 bonds that connect the emptied top-left site with its nearest neighbors.

\begin{figure}[t!]
\includegraphics[width=0.9999\columnwidth]{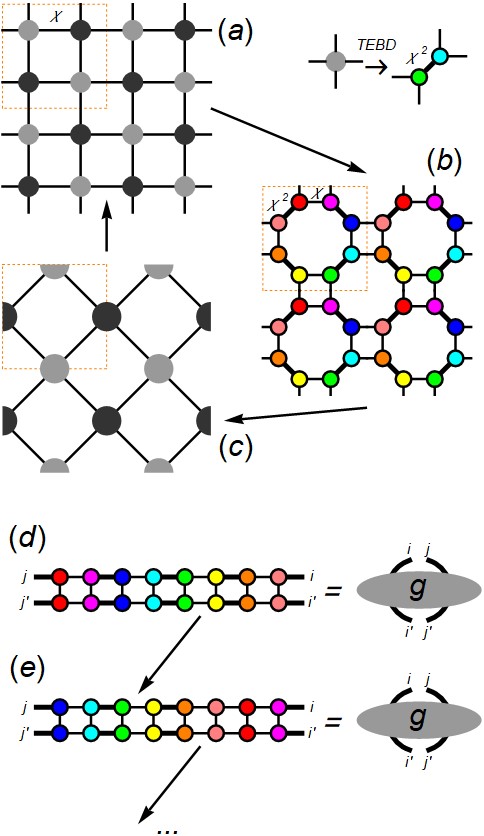}
\caption{
{\bf TRG and zero mode truncation. }
In (a,b,c) one TRG iteration step is summarized. In the move from (a) to (b) each rank-4 tensor is singular value decomposed into a pair of rank-3 tensors creating new bonds with dimension $\chi^2$.
The new bonds are compressed back to $\chi$ in a way that minimizes the Frobenius norm between the rank-8 tensors enclosed by the dashed orange rectangles in (a) and (b). The rectangle in (a) is an 8-index target state and the rectangle in (b) contains 8 rank-3 variational tensors.
After the compression, in the move from (b) to (c) groups of four rank-3 tensors are contracted into new coarse-grained tensors. Then the whole (a,b,c)-iteration is repeated with the new tensors.
In (d) a metric tensor used for zero mode truncation of the first (pink-red) $\chi^2$-bond. After this bond is truncated to $\chi$, the second (magenta-blue) $\chi^2$-bond is truncated with the metric tensor in (e), and so on until truncation of all four $\chi^2$-bonds.
}\label{fig:TRG}
\end{figure}

\begin{figure}[t!]
\includegraphics[width=0.9999\columnwidth]{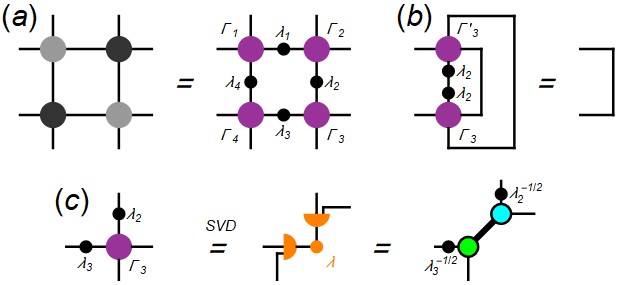}
\caption{
{\bf TRG: from a rank-4 to two rank-3 tensors. }
In (a) the periodic MPS is brought to the Vidal gauge with site tensors $\Gamma_i$ and diagonal bond tensors $\lambda_i$.
They satisfy orthogonality relations whose example is shown in (b), where $\Gamma'_3$ is a mirror image of $\Gamma_3$ with respect to the horizontal axis.
Panel (c) shows how $\Gamma_3$ is decomposed into two rank-3 tensors. $\Gamma_3$, together with its adjacent $\lambda_2$ and $\lambda_3$, is singular value decomposed (SVD) into two isometries and a diagonal matrix of singular values $\lambda$. After $\lambda=\lambda^{1/2}\lambda^{1/2}$ is symmetrically absorbed into the isometries the rank-3 tensors are obtained.
}\label{fig:BP}
\end{figure}

\begin{figure}[t!]
\includegraphics[width=0.9999\columnwidth]{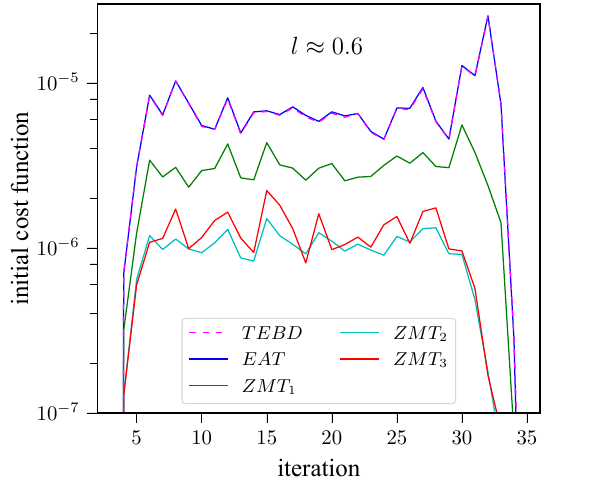}
\caption{
{\bf TRG: initial cost function. }
The cost function right after the initialization in Fig. \ref{fig:TRG} (d,e,...) in function of the iteration number. Five truncations schemes are shown.
TEBD is the simple (and standard) $\lambda$-truncation. Its accuracy is essentially the same as EAT, as both schemes ignore the loopiness of the periodic MPS.
The scheme ZMT$_1$ with diagonal $Z_{ij}=Z_i\delta_{ij}$ is better than TEBD/EAT.
ZMT$_2$ with symmetric $Z_{ij}=Z_{ji}$ is generally the best.
ZMT$_3$ with unrestricted $Z_{ij}$ is typically worse than the symmetric ZMT$_2$.
Here the bond dimension $\chi=16$ and, in ZMT$_{2,3}$, the threshold $\delta=10^{-10}$.
The loopiness \eqref{eq:l} is $l\approx 0.6$.
}\label{fig:cf}
\end{figure}

\section{Tensor renormalization group}
\label{sec:TRG}

The classical Ising model on an infinite square lattice is described by
\be
H=-\sum_{\langle i,j \rangle} \sigma_i\sigma_j,
\ee
where $\sigma_i=\pm1$. As shown by Onsager, the partition function, $Z={\rm Tr}e^{-\beta H}$, can be expressed as tensor-trace over a tensor network on a square lattice. The tensor elements are given by
\bea
&& T_{1,2,1,2}=T_{2,1,2,1}=e^{-4\beta}, \nonumber\\
&& T_{1,1,1,1}=T_{2,2,2,2}=e^{4\beta},
\eea
and $1$ otherwise. The tensor renormalization group (TRG) \cite{TRG_entaglement_filtering,TRG_loop_renormalization,TRG_Nuclear_Norm} is outlined in Fig. \ref{fig:TRG} (a,b,c). It is a way to coarse grain the tensor network so much that the partition function can be obtained efficiently and accurately as a tensor-trace over a few lattice sites. The bond dimension of the coarse grained tensors is limited by $\chi$ that is a refinement parameter.

The four tensors enclosed by the dashed orange rectangle in Fig. \ref{fig:TRG} (a) are interpreted as a periodic matrix product state (pMPS) whose ``physical'' indices are the lines sticking out of the rectangle. The pMPS is brought to the Vidal gauge with site tensors $\Gamma_i$ and bond matrices $\lambda_i$, see Fig. \ref{fig:BP} (a). In the move from (a) to (b), the rank-4 tensor is decomposed into pairs of rank-3 tensors by SVD of the site tensor including its neighboring bond matrices, see \ref{fig:BP} (c). As a result, new bonds appear with dimension $\chi^2$ and the eight rank-3 tensors enclosed by the dashed orange rectangle in Fig. \ref{fig:TRG} (b) make a new pMPS. Its $\chi^2$-bonds have to be compressed down to $\chi$. The simplest compression is to truncate the singular values $\lambda$ in \ref{fig:BP} (c). This local compression takes into account only the local environment of the truncated bond including the rank-4 tensor $\Gamma_i$ and its adjacent bond matrices $\lambda_{i-1}$ and $\lambda_{i}$. It cannot see that the pMPS is periodic.

Once all the $\chi^2$-bonds are truncated one way or another, the truncated rank-3 tensors are optimized to minimize the cost function equal to the Frobenius norm squared of the difference between the two pMPS's in Figs. \ref{fig:TRG} (a) and (b). Here (a) is the target pMPS and (b) is a contraction of 8 variational rank-3 tensors. The cost function is a paraboloid in each of the variational tensors when the other are fixed. Therefore, they can be quickly optimized one-by-one in a loop repeated until convergence of the cost function.

We compare the simple $\lambda$-truncation of the variational tensors with three variants of the zero mode truncation. The initial truncation proceeds in 4 steps shown in Fig. \ref{fig:TRG} (d,e,...). In (d) the bond metric tensor \eqref{eq:g_ij} is defined for the first $\chi^2$-bond. In ZMT$_1$ we use the more compact metric $g_{ij}=g_{ii,jj}$ in \eqref{eq:g_ij} and truncate as in \eqref{eq:overlap_truncation}. The dimension is cut one-by-one all the way down to $\chi$. In ZMT$_2$ the one-by-one ZMT$_1$ truncation proceeds down to $\chi'>\chi$ when the error in \eqref{eq:fD} becomes greater than a threshold $\delta$ for the first time. From $\chi'$ on it continues as in Sec. \ref{sec:general} but with eigenmodes restricted to the real symmetric subspace, $Z_{ij}=Z_{ji}$, where eigenvalues $E_k$ and eigenvectors $S_{kj}$ are real. The third variant, ZMT$_3$, is the same as ZMT$_2$ but $Z_{ij}$ is not assumed symmetric. Instead $E_D$ is chosen to be the largest real eigenvalue of $Z_{ij}$ (with real eigevectors).

Fig. \ref{fig:cf} shows the initial cost function right after the initialization in Fig. \ref{fig:TRG} (d,e,...). The simple $\lambda$-truncation (TEBD) and EAT are essentially the same, as they both ignore loopiness of the periodic MPS. ZMT$_{1}$ with diagonal $Z_{ij}$ is better and ZMT$_{2,3}$ with symmetric/unrestricted $Z_{ij}$ are the best. Interestingly, the symmetric ZMT$_{2}$ is typically better than ZMT$_{3}$, where $Z_{ij}$ itself is unrestricted but one is forced to choose among its real eigenvalues $E_k$ only.

\section{Conclusion}
\label{sec:concl}

The series of examples demonstrates that the zero-mode truncation --- that can be also interpreted as the zero-mode gauge fixing or elimination of linear dependence --- in general provides better initial truncation than methods that ignore loopiness of the tensor network. The initialization is followed by further variational optimization. In order to prevent it from trapping in a local minimum, it is essential to provide good enough initialization, as we can see in the example with $Z_2$ gauge fields. The good initialization can also become essential for more complex problems treated by symmetric tensor networks, where the initial sizes of symmetry sectors remain unchanged in the following optimization.

The data used for the figures in this article are openly available from the RODBUK repository at Ref.~\onlinecite{UJ/0HZFA7_2025}.

\acknowledgments

We are indebted to Marek Rams for stimulating discussions.
This research was
funded by the National Science Centre (NCN), Poland, under project 2021/03/Y/ST2/00184 within the QuantERA II Programme that has received funding from the European Union's Horizon 2020 research and innovation programme under Grant Agreement No 101017733 (I.S.),
funded by the National Science Centre (NCN), Poland, under projects 2024/55/B/ST3/00626 (J.D.) 
and 2019/35/B/ST3/01028 (Y.Z.). 
%
%
This research was also supported by a grant from the Priority Research Area DigiWorld under the Strategic Programme Excellence Initiative at Jagiellonian University (JD).

\bibliographystyle{apsrev4-2}
\bibliography{KZref.bib}

\appendix

\section{Imperfect linear dependence}
\label{app:imperfect_ld}

When the zero mode is not perfect, and $N_D>0$ in \eqref{eq:g_i}, then the truncated state \eqref{eq:overlap_truncation} can still be considered as a variational state:
\be
\ket{\psi_z} =
\sum_{j=1}^{D-1} \left(1+z Z_j\right) \ket{\psi_j}.
\label{eq:psi_z_app}
\ee
Here $z$ is a variational parameter. The norm squared of the difference between the variational state and the target state, $\ket{\psi} = \sum_{j=1}^{D} \ket{\psi_j}$, equals
\bea
f &=& N_D |Z_D|^{-2} w^*w + g_{DD} (1+w^*)(1+w) \nonumber \\
  & & - N_D w^* (1+w) - N_D w (1+w^*).
\eea
Here $w=Z_Dz$, is a variational parameter.
If we keep
\be
w_0=-1,
\ee
equivalent to $z_0=-1/E_D$ that would be optimal for an exact zero mode, then the norm's value is
\be
f_0 = N_D|Z_D|^{-2}.
\label{eq:fDD_ld}
\ee
Rather than the lowest eigenmode with the minimal eigenvalue $N_D$, it is better to choose the one with the lowest $f_0$. Both $N_D$ and its maximal $|Z_D|$ matter.

More accurately, the norm is minimal at
\be
w_{\rm min} = w_0~ \frac{1-n}{1-n(2-|Z_D|^{-2})}
\label{eq:wmin_ld}
\ee
when it is equal to
\be
f_{\rm min} =
f_0~
\frac{1-n|Z_D|^{2}}{1-n(2-|Z_D|^{-2})}
\ee
Here $n=N_D/g_{DD}={\cal O}(N_D)$.
Formally, to leading order in small $N_D$, we have $w_{\rm min}\approx w_0$ and $f_{\rm min}\approx f_0$ used in the main text.

\section{Imperfect general zero mode}
\label{app:imperfect}

Suppose that $Z$ is not an exact zero mode, i.e. $g~Z=N~Z$ with $N>0$, but we still consider \eqref{eq:psi_z_gen} as a variational state. The small non-zero $N$ results in a finite truncation error. The state \eqref{eq:psi_z_gen} can be also written as
\be
\ket{\psi'}=
\sum_{k=1}^{D-1}
\left( 1 + E_k z \right)
\ket{\phi_k}.
\label{eq:psi'}
\ee
Here we define states
\be
\ket{\phi_k} =  \sum_{i,j=1}^D S^{-1}_{ik} S_{kj} \ket{\psi_{ij}}.
\ee
In their terms the exact target state \eqref{eq:psi_ij} becomes
\be
\ket{\psi}=\sum_{k=1}^{D}\ket{\phi_k}.
\label{eq:psi_target}
\ee
The norm squared of the difference between \eqref{eq:psi'} and \eqref{eq:psi_target} is
\be
f = N A ~ w^*w - 2N {\rm Re}~ B~ w^* (1+w) + C (1+w^*)(1+w) .
\ee
Here $w=E_D z$ is a variational parameter and the coefficients are
\bea
A &=& |E_D|^{-2}, \\ 
B &=&  E_D^{-1} \sum_{ij} Z^*_{ij} \left( S^{-1}_{iD} S_{Dj} \right), \\
C &=& \sum_{iji'j'}  \left( S^{-1}_{iD} S_{Dj} \right)^*
                            g_{ij,i'j'}
                     \left( S^{-1}_{i'D} S_{Dj'} \right).
\eea
When we keep
\be
w_0=-1,
\ee
equivalent to $z_0=-1/E_D$ that would be optimal for an exact zero mode, then the norm's value is
\be
f_0 = N A = N|E_D|^{-2}.
\label{eq:fDD}
\ee
Rather than the eigenmode with the minimal eigenvalue $N$, it is better to choose the one with the lowest $f_0$. Both $N$ and its maximal $|E_D|$ matter.
More accurately, the norm is minimal at
\be
w_{\rm min} = w_0~ \frac{1-nB}{1-n(B+B^*-A)}
\label{eq:wmin}
\ee
when it is equal to
\be
f_{\rm min} = f_0~ \frac{1-n(BB^*/A)}{1-n(B+B^*-A)}.
\ee
Here $n=N/C={\cal O}(N)$.
Formally, to leading order in the small $N$, we have $w_{\rm min}\approx w_0$ and $f_{\rm min}\approx f_0$ used in the main text.

\section{Improving zero mode for truncation}
\label{app:beyond}

Up to this point we employed the eigenmodes of the metric $g_{ab,cd}$ as means for truncation. The optimal eigenmode $Z_{ab}$ was selected as the one with the minimal truncation error \eqref{eq:fDDmin}. In this Appendix we modify the optimal eigenmode from $Z_{ab}$ to ${\cal Z}_{ab}$ to lower its truncation error even further. To this end we write \eqref{eq:fDDmin} as
\be
f=\lim_{n\to\infty}
       \frac{ \cal{N} }
            { \left( {\rm Tr }{\cal Z}^n {\rm Tr }{\cal Z}^{\dag n} \right)^{1/n} }
            \label{eq:limf}
\ee
Here ${\cal N}=\sum_{abcd} {\cal Z}^{*}_{ab} g_{ab,cd} {\cal Z}_{cd}$ and we assume the usual normalization ${\rm Tr} {\cal Z}^\dag {\cal Z} =1$. For the optimal ${\cal Z}$ the gradient is zero:
\be
P_{ab,cd} ~ \frac{\partial f}{\partial {\cal Z}^*_{cd}}=0.
\label{eq:Pgrad}
\ee
Here the projector
$
P_{ab,cd} = \delta_{ab,cd} - {\cal Z}_{ab} {\cal Z}^*_{cd}
$
takes care of the normalization constraint. The equation \eqref{eq:Pgrad} reads
\bea
&&
\left( g_{ab,cd} - {\cal N} \delta_{ab,cd} \right) {\cal Z}_{cd} =
{\cal N}
\left(
\lim_{n\to\infty}
\frac{{\cal Z}^{\dag(n-1)}_{ab}}{{\cal E}_D^{*n}} - {\cal Z}_{ab}
\right).
\eea
Here $\cal{E}_D$ is the largest-magnitude eigenvalue of ${\cal Z}_{ab}$.

We assume ${\cal Z}$ to be the optimal eigenmode $Z$ plus a perturbation $\varepsilon$:
\be
{\cal Z}=Z+f~\varepsilon.
\label{eq:Zf}
\ee
Here $f$ is the small truncation error for the eigenmode \eqref{eq:fDDmin} and the perturbation is orthogonal to the eigenmode: ${\rm Tr} Z^\dag\varepsilon=0$. To leading order in $\varepsilon$ the perturbation satisfies:
\be
\left( g_{ab,cd} - N \delta_{ab,cd} \right) \varepsilon_{cd} =
\left|E_D\right|^2
\left( \lim_{n\to\infty} \frac{Z^{\dag(n-1)}_{ab}}{E_D^{*n}} - Z_{ab} \right)
\ee
With the eigendecomposition \eqref{eq:Zdiag}, after taking the limit $n\to\infty$, we obtain
\bea
&&
\left( g_{ab,cd} - N \delta_{ab,cd} \right) \varepsilon_{cd} = \nonumber\\
&&
\left(
S^\dag_{aD} E_D S^{-1\dag}_{Db}  - \left|E_D\right|^2 \sum_k S^{-1}_{ak} E_k S_{kb}
\right)
\eea
The matrix $g_{ab,cd} - N \delta_{ab,cd}$ is singular. A solution by pseudoinverse yields $\varepsilon$ that is orthogonal to $Z$.

The equation may be more transparent for a unitary $S_{kb}=U_{kb}$ and real $E_D=E_D^*$:
\bea
&&
\left( g_{ab,cd} - N \delta_{ab,cd} \right) \varepsilon_{cd} =  \\
&&
\left(1-\left|E_D\right|^2\right) U^*_{aD} E_D U_{Db}
- \left|E_D\right|^2 \sum_{k\neq D} U^*_{ak} E_k U_{kb}.        \nonumber
\eea
In the total ${\cal Z}$ in \eqref{eq:Zf} the solution $\varepsilon$ enhances the leading eigenvalue $E_D$ and suppresses the subleading $E_{k\neq D}$.

The perturbative solution \eqref{eq:Zf} increases both ${\cal N}>N$ and $|{\cal E}_D|>|E_D|$ but in such a way that the truncation error \eqref{eq:limf} becomes smaller than the zero-mode error $f$ in \eqref{eq:fDDmin}. However, the improvement is formally only ${\cal O}(f^2)$ and this estimate is corroborated by numerical tests.


\end{document}